\documentclass[pra,superscriptaddress, nofootinbib, twocolumn]{revtex4}
\usepackage{graphicx}       
\usepackage{bm}         
\usepackage{amsfonts}
\usepackage{amsmath}        
\usepackage{latexsym}
\usepackage{amssymb}
\usepackage{color}

\begin{document}

\title{The information of high-dimensional time-bin encoded photons}
\author{Thomas Brougham$^1$, Christoph F. Wildfeuer$^{2}$, Stephen M. Barnett$^1$ and  Daniel J. Gauthier$^{3,4}$\\ 
$^1$ School of Physics, University of Glasgow, Glasgow, G12 8QQ, U.K.\\
$^2$ Institute of Mathematics and Natural Sciences, University of Applied Sciences and Arts, North Western, Switzerland, Bahnhofstrasse 6, 5210 Windisch, Switzerland \\
$^3$ Department of Physics, Duke University, Box 90305, Durham, North Carolina 27708, USA\\
$^4$ Department of Physics, Ohio State University, Columbus, Ohio 43210, USA\\}

\begin{abstract}

\par High-dimensional entanglement is an important physical resource for quantum communication.  A basic issue for any communication scheme is how many shared bits two two parties can extract subject to experimental noise.  We determine the shared information that can be extracted from time-bin entangled photons using frame encoding.  We consider photons generated by a general down-conversion source and also model losses, dark counts and the effects of multiple photons within each frame.  Furthermore, we describe a procedure for including other imperfections such as after-pulsing, detector dead-times and jitter.  The results are illustrated by deriving analytic expressions for the maximum information that can be extracted from {\it high-dimensional} time-bin entangled photons generated by a spontaneous parametric down conversion.  A key finding is that under realistic conditions and using standard SPAD detectors one can still choose frame size so as to extract {\it over} 10 bits per photon.  These results are thus useful for experiments on high-dimensional quantum-key distribution system, but are not limited to such systems.  For example, the results are also useful in determining the limits of fibre arrays or within time-multiplexing schemes.

\end{abstract}
\maketitle

\section{Introduction}
\par It is well know that entangled photons can be used to extract shared random bits.  
The number of extractable bits per photon pair depends on the dimensions of the entangled degree of freedom.  For example, polarization entanglement allows at most one shared bit per photon pair.  An alternative is to use the arrival time of a photon.  Encoding within the arrival time of a pair of photons offers an experimentally viable way of generating high-dimensional entangled states \cite{etime,tb1,tb2,tb3}.  High-dimensional entangled states have many interesting properties \cite{NRA,CBKG,jon} and can allow for multiple shared bits extracted from each photon pair.  This can be beneficial for quantum key distribution (QKD), where each detected photon pair could encode over 10 bits of information \cite{Ali-Khan}.

There are several benefits to encoding within the time of arrival as opposed to other degrees of freedom, such as the spatial modes.  One key advantage is to minimize the effects of detector dead-time, which not only limits the rate at which information can be communicated, but also impacts on security within QKD \cite{deadtimeattack,deadtimeattack2}.  Another benefit is that temporal modes can be easily coupled into fibres, which is not the case for beams with non-zero orbital angular momentum \cite{Mair,gotte,leach}.

It is clear that imperfections such as loss have a strong effect on how much information we can extract from high-dimensional entangled photons.  It is thus vital to model the effects of realistic experimental errors.  Any model must take account of the photons source, channel losses and imperfect detectors.  Nevertheless, it has been shown that it may still be possible to extract over 10 bits per photon pair under reasonable experimental conditions \cite{Thomas}.  

In practice, the amount of extractable information depends on the error correcting scheme.  In turn, this can depend on the physical implementation.  The case of time-bin encoding raises specific problems.  For example, in standard polarization based QKD schemes, one uses the timing information to help correct losses.  It is thus possible to remove all the cases where Alice and Bob do not share coincident photons.  
This approach is clearly not suitable when information is encoded in the arrival time.  Instead, we require a method for correcting errors that does not reveal the timing information.  A common way of circumventing this is to split the arrival time into time-bins, which are then grouped together to form a {\it frame} \cite{Kochman}.  Alice and Bob then publicly announce the number of photons detected in each frame.  The use of frame encoding, while greatly facilitating error correction, does add an additional constraint to the extractable information.  A realistic model must take this into account.


The aim of this work is to determine the extractable information from high-dimensional, temporal-entangled photons.  In particular, we determine the maximum number of shared bits that, on average, one obtains using frame encoding schemes.  It is important to stress that while the main motivation for this work comes from QKD, we are {\it not} proposing a new QKD protocol.  As such we do not concern ourselves with the task of securing the bits.  Instead, we establish the maximum shared information that can be obtained via reconciliation.  The task of securing the bits will generally depend on the exact nature of the setup.

The results we present are not only useful for QKD.  For instance, it is has been argued that the mutual information can be used to quantify the entanglement within and SPDC source \cite{entanglementSPDC}.  Furthermore, the results can also be used to quantify the capacity of fibre array \cite{Thomas}, which can be used, for instance, in time-multiplexing of detectors \cite{timemulti1,timemulti2}.

The general formalism we present can model experiments such as illuminating a nonlinear crystal with a mode-locked laser,(see Fig \ref{experiment} and the description in the next section).  The approach is, however, not tied to this setup and applies to general sources of entangled photons.  For instance, the formalism can be applied to cases where the Poissonian approximation is not appropriate.  One could thus use our approach to model many different time-bin based experiments.  In addition, the formalism also takes account of asymmetric channel losses, dark counts, jitter and other such effects.  The breadth and generality of the considered errors is beyond that which is considered in previous works \cite{Kochman,wornell}.

To understand how these results can be useful, consider a QKD experiment with detector jitter.  A common approach to reducing jitter is to increase the width of the time-bins.  This is, however, not always possible or practical.  Furthermore, even when we can increase the time-bin width, this affects the amount of information one can extract.  In this context, an important question is whether it is better to increase the width of the time-bins or to correct the jitter errors using a reconciliation protocol.  To answer this question one must calculate the mutual information in the presence of jitter.  The can be achieved using the results of this paper.

Another way in which our approach goes beyond existing results, such as \cite{Kochman,wornell}, is to calculate explicitly the effect of frames that contain two or more photon pairs.  Our findings are thus complementary to those of \cite{wornell2}, which presents a layered protocol for extracting information from general multi-array frames. The aim of the current work is to find the maximum possible extractable information using any frame-encoding protocol.  This should prove important for optimization and design of new error-correction codes for high-dimensional QKD.  

\section{Frame encoding}
\label{secII}

\begin{figure}[htb] 
\begin{center}\includegraphics[scale=0.35]{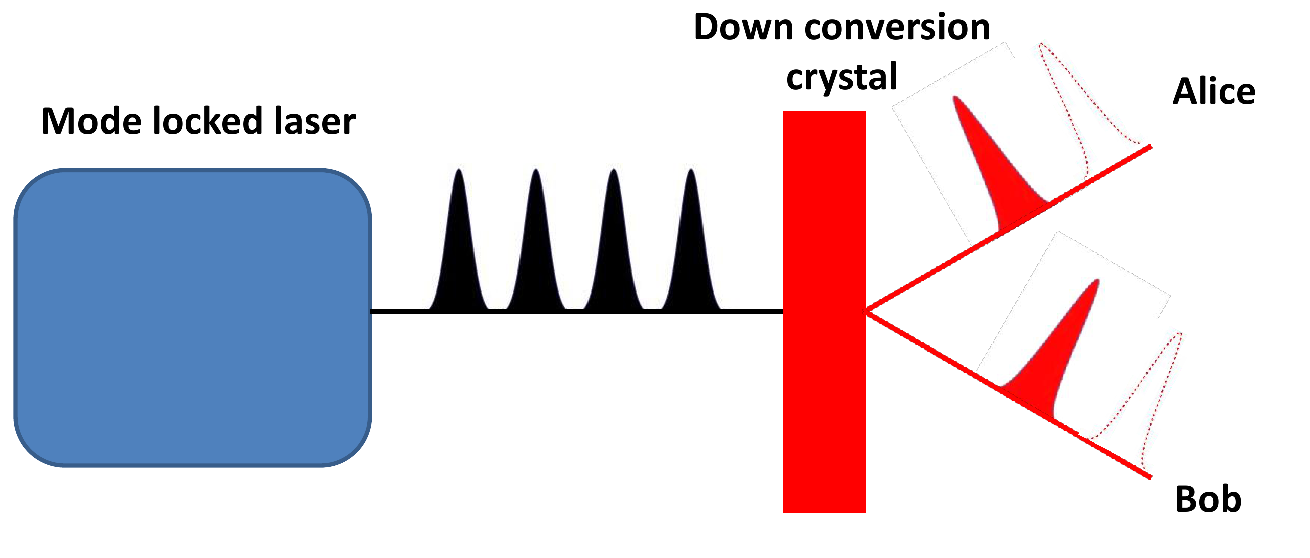}\end{center}%
\caption{\label{experiment}A schematic for a experimental setup that generates and distributes high-dimensional, time-of-arrival entangled photons.  A mode-locked laser generates a coherent train of pulses.  The pulses pumps a nonlinear crystal that produces entangled photon pairs in some of the time-slots; many of the slots contain no photons (color online). }
\end{figure}

Pairs of photons have been prepared experimentally where their arrival time is entangled \cite{etime}.    A common way of generating such photons pairs is to use spontaneous parametric down-conversion (SPDC) \cite{Ali-Khan,Kwiat,Liang}.  Figure \ref{experiment} shows a typical setup, where a nonlinear crystal is pumped by a mode-locked laser \cite{cqo,cqo2}.  The incoming pulses are classically coherent.  As down-conversion is a unitary process, the coherence between the pulses is transferred to a coherence between the amplitudes to generate photon pairs in each time-bin.  In the ideal case, two parties, called Alice and Bob, use this setup to generate a random sequence photon pairs that are perfectly correlated in time. Alice and Bob then use single-photon counter modules and synchronized time-tagging devices to obtain the timing information.  Setups such as this have been realized experimentally \cite{cqo,cqo2,cqo3}. 

To make use of such time-entangled states, the arrival time is divided into a discrete set of time-bins.  For the case of a mode-locked train of pulses, the time-bin width is set by the pulse spacing.  In alternate setups where photons are generated by a single pulse or a continuous-wave laser, the time is discretized by dividing the time into discrete time-bins.  If the widths of these time-bins are chosen appropriately, then Alice and Bob should detect their photons within the same time-bin.  The uncertainty in the arrival time can then be used to extract shared random bits.  An eavesdropper could then be detected by measuring within another basis \cite{cavity,cavity2,dispersion1,dispersion2,dispersion3,franson1,franson2,franson3}. 

In real experiments, there will always be errors.  Alice and Bob will thus carry out error-correction to obtain a shared random string.  A simple approach is to group together several contiguous time-bins to form a frame \cite{Kochman}.  For each frame, Alice and Bob announce the number of time-bins in which they detect photons.  Let $K_A$ and $K_B$ denote the number of time-bins in which photons were detected by Alice and Bob respectively.\footnote{Note that $K_A$ and $K_B$ do not correspond to the number of photons in Alice and Bob's frames.  For example, $K_A=1$ means that Alice detected a click within one time-bin.  In principle, it is possible that this click corresponded to multiple photons or even a dark count.} One can use $K_A$ and $K_B$ to classify the frames; we thus write ($K_A$, $K_B$)-frames, to denote the class of frames where Alice see $K_A$ clicks while Bob sees $K_B$.  Error-correcting codes can then be developed to deal with each class of frame.  In many setups, the chance that Alice and Bob will detect multiple photons within a frame is low.  In such situations, it is sufficient to consider only cases such as $(1,1)$ and $(2,2)$-frames.

We envisage more complicated frame-encoding schemes, where Alice and Bob don't publicly announce $K_A$ and $K_B$.  In all that follows, we consider only simple schemes where $K_A$ and $K_B$ are announced.  The reason is that an understanding of this situation is vital also for the more complicated protocols.  In particular, we will show that the shared information for the case where $K_A$ and $K_B$ are announced differ from the more complicated protocols by a single term.  Thus, the results we present can also be used to calculate the shared information for more advanced protocols.
   
Let $N$ represent the number of time-bins that comprise each frame.  If Alice observes clicks in $K_A$ time-bins, then the number of possible distinct measurement records she could have is given by the Binomial coefficient 
\begin{equation}
\left(\begin{array}{c} N\\ K_A\end{array}\right)=\frac{N!}{K_A!(N-K_A)!}.
\end{equation}
A particular record of measurement results, or {\it measurement patterns}, can be denoted by an $N$-bit binary string that indicate the location of the time-bins where photons are detected.  For example, if $N=4$, $K_A=2$, and Alice sees clicks in the first and third time-bins, then the corresponding binary string is $1010$.  It will prove useful to introduce a further piece of notation.  We denote Alice's measurement patterns symbolically as $A^{K_A}_r$, where $r$ is the binary string that uniquely describes the pattern.  We describe Bob's measurement pattern using the same notation, where each $A$ is changed to $B$.

If there are no errors, then Alice and Bob each see the same measurement pattern.  When losses are present, then Alice and Bob's measurement patterns can be different.  Nevertheless, there will still be some correlation in their results.  We expect that it should still be common to observe clicks within the same time-bins.  For a particular frame, let $L$ be the number time-bins in which they both share clicks.  For example, if Alice has the pattern 0101, while Bob sees the pattern 1100, then $L=1$.  Clearly, $\text{Min}\{K_A,K_B\}\ge L$.  Furthermore, the allowed values for $L$ satisfy the inequality $N\ge K_A+K_B-L$.  For fixed values of $K_A$, $K_B$ and $L$, the total number of different joint patterns for Alice and Bob is $\Omega_L(K_A,K_B)=N!/[L!(K_A-L)!(K_B-L)!(N-K_A-K_B+L)!]$
which is a multinomial coefficient.  
These observations will prove useful later on.


We want to determine the information contained within an {\it average} frame.  The mutual information per frame is denoted as $H_{frame}(A:B)$.  To find the number of shared bits per photon, we divide $H_{frame}(A:B)$ by the average number of photon pairs found within a frame.  We then calculate $H_{frame}(A:B)$ using the method outlined in \cite{Thomas}.  For a frame encoding scheme, we will not reach the bits per photon limit set by $H_{frame}(A:B)$.  The reason for this is that, in a frame encoding scheme, Alice and Bob publicly announce the number of clicks they see in each frame.  They then apply error correction codes {\it individually} to each class of frame.  This necessarily results in a loss of randomness, and hence, of random bits.

The shared information that we can extract is related to the mutual information.  However, it is not $H_{frame}(A:B)$, but instead the mutual information post-selected on when $K_A$ and $K_B$ have specific values.  This implies that we must use conditional probabilities in place of the standard probabilities to determine the conditional mutual information \cite{EIT}. 

Suppose Alice observes $x$ clicks while Bob sees $y$, the maximum shared information per frame is given by the conditional mutual information
\begin{eqnarray}
\label{cond_mi_def}
& H(A:B|K_A=x,K_B=y)=\\
& -\sum_{r,s} P(A^x_r,B^y_s|K_A=x,K_B=y)\nonumber\\
& \times \log_2\left[\frac{P(A^x_r,B^y_s|K_A=x,K_B=y)}{P(A^x_r|K_A=x)P(B^y_s|K_B=y)}\right],\nonumber
\end{eqnarray}
where $P(A^x_r,B^y_s|K_A=x,K_B=y)$ is the joint conditional probability for Alice and Bob to obtain the patterns $A^x_r$ and $B^y_s$, while $P(A^x_r|K_A=x)$ and $P(B^y_s|K_B=y)$ are the marginal conditional probabilities for Alice and Bob, respectively.  The conditional mutual information $H(A:B|K_A=x,K_B=y)$ gives the {\it maximum} number of bits per frame that can be extracted from $(x,y)$-frames.  

One average, Alice and Bob can extract $H(A:B|K_A,K_B)=\sum_{x,y}P(K_A=x,K_B=y)H(A:B|K_A=x,K_B=y)$ bits per frame, where $P(K_A=x,K_B=y)$ is the probability for Alice and Bob to observe $x$ and $y$ clicks, respectively.  Notice that $H(A:B|K_A,K_B)\ne H_{frame}(A:B)$, hence we have lost some information.  We find that $H_{frame}(A:B)=H(A:B|K_A,K_B)+H(K_A,K_B)$, where $H(K_A,K_B)=-\sum_{x,y}{P(K_A=x,K_B=y)\log_2 P(K_A=x,K_B=y)}$, {\it i.e.}, it is the entropy in the uncertainty in the number of clicks per frame.  The loss of information thus follows simply from the fact that Alice and Bob announce the values of $K_A$ and $K_B$.

In a practical application, one may not be able to develop effective error correcting codes for all of the classes of frame.  In this instance, $H(A:B|K_A,K_B)$ will over-estimate the extractable information.  The actual extractable information can be found by averaging $H(A:B|K_A=x,K_B=y)$ over the frames for which we do have error correcting codes.  For example, if we only have codes for (1,1)-frames, then the extractable information is $P(K_A=1,K_B=1)H(A:B|K_A=1,K_B=1)$.

\section{Calculating the conditional probabilities and conditional mutual information}
In this section, we present a general procedure for calculating the conditional probabilities and hence the conditional mutual information.  The approach allows us to calculate the probabilities for a general source and includes the effects of errors such as channel losses and detector imperfections.  The first step is to work out the detection probabilities for a single time-bin.  We then use these to construct the probabilities to observe specific measurement patterns, from which the conditional probabilities are calculated.  

One thing to notice is that the coherence between the time-bins does not appear in our calculations.  The reason for this is that our present results are for the case when we measure within the time-of-arrival basis.  Such measurements cannot detect coherence between time-bins. Furthermore, they will actually destroy any coherence.  The situation we thus consider is one where a measurement has been made of the photons time-of-arrival, which inevitably disturbs the temporal coherence.  The temporal coherence would, however, be reveled if one measured in a basis that was a superposition of time-bins.  For instance, consider a mode-locked laser generating a train of coherence pulses acting as a pump for a non-linear crystal.  The coherence between the time-bins of the down-converted photons is not evident if we measure the time-of-arrival.  However, it has been demonstrated experimentally using a Franson interferometer \cite{kevin}.

\subsection{Single time-bin probabilities}
To calculate the single time-bin detection probabilities, we must first model the source, channel and detectors.  The approach we use is based on that presented in \cite{Thomas}.  We thus only give a brief recap of the important points.  First, we assume that the source produces pairs of entangled photons, where the probability to produce $m$ pairs within any given time-bin is $P_s(m)$.  For simplicity, we initially assume that $P_s(m)$ is the same for each time-bin.  Let $\lambda$ be the average number of photon pairs produced per time-bin, hence $\lambda=\sum_m{mP_s(m)}$.  

The information is encoded in the temporal location of the photon pairs, not their number.  For this reason, we assume that the detectors do not resolve photon number.  For ideal detectors, with no losses, the probability to observe a click in a time-bin is $\sum_{m=1}^{\infty}{P_s(m)}$.  All real detectors, however, suffer losses.  
Let Alice and Bob's detectors loss be $\xi_A$ and $\xi_B$ respectively. 
The probability to detect a single photon that is incident on Alice's detector, is thus $\xi_A$.  In addition to the losses due to inefficiencies in the detector, there are also losses from transmission of the photons from the source to the detectors.  Let $\eta_{a}$ and $\eta_{b}$ be the losses in Alice and Bob's channels, respectively.  Combining  the two sources of loss into a single total efficiency, Alice's total efficiency is $\eta_A=\xi_A\eta_{a}$, and Bob's total efficiency is $\eta_B=\xi_A\eta_{b}$.     

We are now in a position to calculate the probability for Alice and Bob to observe photons within a single time-bin.  The key mathematical method is to use moment generating function to include the effects of loss.  See \cite{Barnett} for a full discussion on moment generating functions and their properties.

For a source described by the probability distribution $P_s(m)$ and total losses for Alice and Bob of $\eta_{A}$ and $\eta_B$, respectively, we define the moment generating function
\begin{equation}
\label{momentgf}
M(\nu,\xi)=\sum_{m=0}^{\infty}{P_s(m)(1-\eta_A\nu)^m(1-\eta_B\xi)^m},
\end{equation}
where we have neglected dark counts for now.  Consider a single time-bin.  The probability for Alice and Bob to observe a click within a given time-bin is denoted by $\pi^{AB}_{i,j}$, where $i,j\in\{0,c\}$ and $c$ represents a click while $0$ signifies no click.  It can be shown that the probabilities are \cite{Thomas}
\begin{eqnarray}
\label{piprobs}
\pi^{AB}_{00} &=& M(1, 1), \\
\pi^{AB}_{c0}&=&\sum_{l=1}^\infty \frac{1}{l!}\left. \left(-\frac{d}{d \xi}\right)^l M(1, \xi)\right|_{\xi = 1},\nonumber\\
\pi^{AB}_{0c}&=&\sum_{l=1}^\infty \frac{1}{l!}\left. \left(-\frac{d}{d \nu}\right)^l M(\nu, 1)\right|_{\nu = 1},\nonumber\\
\pi^{AB}_{cc}&=&\sum_{n=1}^\infty P_s(n)\left[1 - (1 - \eta_A)^n\right]\left[1-(1-\eta_B)^n\right].\nonumber
\end{eqnarray} 
The effect of dark counts is taken account of using the following procedure.  Let ${P}_{ij}$ represent Alice and Bob's probability to detect photons within a single time-bin when dark counts are present.  We find that  
\begin{eqnarray}
\label{fullprobs}
{P}_{00} &=& (1 - q)^2\pi^{AB}_{00} \\
{P}_{0c} &=& (1 - q)\pi^{AB}_{0c} + (1-q)q\pi^{AB}_{00} \nonumber \\
{P}_{c0} &=& (1 - q)\pi^{AB}_{c0} + (1-q)q\pi^{AB}_{00}\nonumber \\
{P}_{cc} &=& \pi^{AB}_{cc} + q\pi^{AB}_{0c} +q\pi^{AB}_{c0}+ q^2\pi^{AB}_{00} \,,\nonumber
\end{eqnarray}
where $q$ is the probability to observe a dark count in a single time-bin.  We see that the above probabilities sum to one.  The marginal probabilities $P^A_i$ and $P^B_j$ are found from the joint probability $P_{ij}$.  A key feature of these general expression for ${P}_{ij}$ is that they are valid for any choice for the source probability $P_s(m)$.  Thus, our results are not be limited to any particular physical implementation.

\subsection{Probabilities for each frame}
The probability for Alice (or Bob) to observe a particular measurement pattern is calculated using the relevant single time-bin probabilities $P^A_i$ or $P^B_j$.  The probability for Alice to see a pattern $A^{K_A}_r$ is $P(A^{K_A}_r)=[P^A_c]^{K_A} [P^A_0]^{N-K_A}$.  The total probability for Alice to observe a measurement pattern with photons detected in $x$ time-bins is 
\begin{equation}
\label{pka}
P(K_A=x)=\left(\begin{array}{c} N\\ x\end{array}\right)[P^A_c]^x[P^A_0]^{N-x}.
\end{equation}
The probabilities for Bob have the same form, but instead use the probabilities $P^B_j$.

The joint detection probabilities $P(A^{K_A}_r,B^{K_B}_s)$ is calculated using the single time-bin joint detection probabilities.  We find that  
\begin{eqnarray}
\label{pj}
P(A^{K_A}_r,B^{K_B}_s)=[P_{cc}]^L[P_{c0}]^{K_A-L}\nonumber\\
\times[P_{0c}]^{K_B-L}[P_{00}]^{N-K_A-K_B+L}.
\end{eqnarray}
The total probability for Alice and Bob to detect photons within $x$ and $y$ time-bins, respectively, is thus 
\begin{eqnarray}
\label{pkakb}
P(K_A=x,K_B=y)=\sum_L\Omega_L(K_A,K_B)\nonumber\\
\times[P_{cc}]^L[P_{c0}]^{x-L}[P_{0c}]^{y-L}[P_{00}]^{N-x-y+L},
\end{eqnarray}
where $\Omega_L(K_A,K_B)$ denotes the multinomial coefficient $N!/[L!(K_A-L)!(K_B-L)!(N-K_A-K_B+L)!]$.  Equation (\ref{pkakb}) leads to the same marginal probabilities as in Eq. (\ref{pka}).

The conditional probabilities are calculated from Eqs. (\ref{pka}), (\ref{pj}) and (\ref{pkakb}), by recalling the definition of a conditional probability: $P(X|Y)=P(X,Y)/P(Y)$.  We find that
\begin{eqnarray}
\label{condprobs}
& P(A^x_r|K_A=x)=\left(\begin{array}{c} N\\ x\end{array}\right)^{-1},\\
& P(B^y_s|K_B=y)=\left(\begin{array}{c} N\\ y\end{array}\right)^{-1},\nonumber\\
& P(A^x_r,B^y_s|K_A=x,K_B=y)=\frac{P^L_{cc}P^{x-L}_{c0}P^{y-L}_{0c}P^{N-x-y+L}_{00} }{P(K_A=x,K_B=y)},\nonumber
\end{eqnarray}
where $P(K_A=x,K_B=y)$ is given in Eq. (\ref{pkakb}).  The expressions for $P(A^x_r|K_A=x)$ and $P(B^y_s|K_B=y)$ are not true when $\eta_A=\eta_B=1$.  The reason for this is that, in this limit, Alice and Bob must observe the same patterns and $K_A=K_B$.  This implies that $P(K_A=x)$ and $P(K_B=y)$ will be zero if $x\ne y$.  In the case of $\eta_A=\eta_B=1$, then $K_A=K_B=K$ and the conditional mutual information has the simple form 
\begin{equation}
\label{noloss}
H(A:B|K=x)=\log_2\left(\begin{array}{c} N\\ x\end{array}\right).
\end{equation}
In the remainder of this section, we focus on the case where both $\eta_A$ and $\eta_B\ne 1$.

\subsection{Information per photon pair}
We calculate the various entropic quantities, by using Eqs. (\ref{pka}) through to (\ref{condprobs}).  For instance, we find that $H_{frame}(A:B)=N H(A:B)$, where $H(A:B)$ is the mutual information per time-bin.  Suppose we only have error correcting codes for (1,1)-frames, {\it i.e.}, frames where Alice and Bob both announce that they each observe a single click.  The maximum extractable information is $H_{1,1}(A:B)=P(K_A,K_B)H(A:B|K_A=1,K_B=1)$.  

A typical application of this theory is in high-dimensional QKD, which  aims to encode multiple bits on each photon pair.  It is thus worth considering the bits per photon pair.  The average number of photon pairs generated within each frame is $N\lambda$.  Due to losses, the average number of photon pairs that one {\it detects} per frame is $N(\eta_A\eta_B\lambda+q^2)$.  The average number of bits per generated photon pair is $H_{1,1}(A:B)/(N\lambda)$, while the average number of bits per detected photon pair is $H_{1,1}(A:B)/(N[\eta_A\eta_B\lambda+q^2])$.  These quantities can be calculated using the conditional probabilities given in Eq (\ref{condprobs}), within Eq (\ref{cond_mi_def}).  We find that, by using only $(1,1)$-frames, the average number of shared bits per detected photon is
\begin{eqnarray}
\label{postinfophoton}
H_{d}(A:B|K_A=1,K_B=1)=\frac{P^{N-2}_{00}}{\eta_A\eta_B\lambda+q^2}\Big(\Gamma\log \frac{N}{\Gamma}\nonumber\\
+P_{cc}P_{00}\log(P_{cc}P_{00})+(N-1)P_{c0}P_{0c}\log(P_{c0}P_{0c})\Big),\nonumber\\
\end{eqnarray}
where $\Gamma=(N-1)P_{c0}P_{0c}+P_{cc}P_{00}$ and $P_{ij}$ are given in Eq. (\ref{fullprobs}).  The expression for the average number of bits per generated photon has the same form as (\ref{postinfophoton}), but with $\eta_A\eta_B\lambda+q^2$ replaced in the denominator by $\lambda$.  We stress that Eq. (\ref{postinfophoton}) includes the effects of losses, dark counts and a general source.  Furthermore, it can also be applied to situations where the dark count rates are different on each side.  This is accomplished by modifying only the probabilities $P_{ij}$, not the form of Eq. (\ref{postinfophoton}).  

Suppose we have error-correcting codes that work for (2,2)-frames.  The extractable information when using only these frames is found using Eqs. (\ref{condprobs}) together with (\ref{cond_mi_def}) and (\ref{pkakb}), giving the maximum number of bits per detected photon pair as 
\begin{eqnarray}
\label{post2}
H_{d}(A:B|K_A=2,K_B=2)=\frac{(N-1)P_{00}^{N-2}}{\eta_A\eta_B\lambda+q^2}\Big\{\Omega\nonumber\\
\times\log\left[\frac{N(N-1)}{4\Omega}\right]+(P_{cc}P_{00})^2\log[P_{cc}P_{00}]\nonumber\\
+2(P_{c0}P_{0c})^2\log[P_{c0}P_{0c}]\nonumber\\
+(N-2)P_{cc}P_{c0}P_{0c}P_{00}\log[P_{cc}P_{c0}P_{0c}P_{00}]\Big\},\nonumber\\
\end{eqnarray}
where $\Omega=\frac{1}{2}(P_{00}P_{cc})^2+(N-2)P_{cc}P_{c0}P_{0c}P_{00}+\frac{1}{4}(N-2)(N-3)(P_{c0}P_{0c})^2$.  As with Eq. (\ref{postinfophoton}), we obtain the information per generated photon pair by replacing $\eta_A\eta_B\lambda+q^2$ in the denominator with $\lambda$.

If we have error correcting codes for both (1,1) and (2,2)-frames, then the total amount of extractable information is the sum of the information for the two cases, {\it i.e.}, $H_{d}(A:B|K_A=1,K_B=1)+H_{d}(A:B|K_A=2,K_B=2)$.  In general, the information we can extract using only $(x,y)$-frames is calculated using the conditional probabilities (\ref{pkakb}) and (\ref{condprobs}).  As $x$ and $y$ become large, the resulting expressions become more complex.  Nevertheless, we can still obtain analytic expressions by following the same straightforward procedure.

As Alice and Bob publicly announce $K_A$ and $K_B$, they are losing all the information contained within the correlation of these quantities.  It is possible to develop approaches that retain some of this information.  As showed in Sec. \ref{secII}, the correlations in $K_A$ and $K_B$ give a total contribution of $H(K_A,K_B)$ bits per frame to the total shared information per frame.  In terms of bits per detected photon pair, we can gain an additional $H(K_A,K_B)/(\eta_A\eta_B\lambda+q^2)$ bits per photon.  In practice, protocols generally will access a certain fraction $f$ of these bits, where $0\le f\le 1$.  

\section{Additional errors: after-pulsing and detector dead-time}
\label{secIV}
Losses and dark counts are not the only errors that affect the extractable information.  There are also effects such as detector jitter, after-pulsing and detector dead-times.  The discussion of jitter is postponed until the next section.  In this section, we explain how the formalism is modified to describe after-pulsing and dead-time.  

After-pulsing occurs when the detection of a photon sets up a feedback process that can lead to the detector registering a click at a later time \cite{fang}.  After-pulsing will thus temporarily increases our chance to see a dark count after we register a click.  One approximate  model of after-pulsing is to increase the dark count probability $q$ for some fixed number of time-bins $\beta$ after a detection.  
One important feature of after-pulsing is that it occurs regardless of what triggered the detector.  This means that after-pulsing occurs also for dark counts.  The single time-bin detection probabilities, Eqs. (\ref{fullprobs}), include contributions from dark counts.  This means that our approach will take account of after-pulsing that is generated both by photons and from dark counts.

The value for $\beta$ can be large \cite{fang}.  This means that a click near the beginning of a frame can result from after-pulsing from the previous frame.  Similarly, the average position of detected photons is random, which means the location of the $\beta$ time-bins will also be random.  Recall, however, that we are calculating the shared information for an average frame.  To take account of these difficulties, the fairest approach is to modify $q$ for all time-slots.  In this case, the information per photon will retain the form given in in Eqs. (\ref{postinfophoton}) and (\ref{post2}), but where the value of $q$ has been suitably increased.

After a photon is detected, it is common for a detector to loose sensitivity to subsequent photons for a period of time.  This interval of time is know as the detector's {\it dead-time} \cite{Bedard}.  If the duration of the dead-time is equal to the width of $M_d$ time-bins, then we will not observe photons for at least the next $M_d$ time-slots after a detection.  Dead-time is not a serious problem for (1,1)-frames, provided that the frame is longer than the period of dead-time.  In this limit, Eq. (\ref{postinfophoton}) is still valid.  However, the effects of detector dead-time will be important for classes of frames such as (2,2)-frames.  For these cases, we must adopt the following modified procedure.  

First, we calculate the moment generating function and the single time-bin probabilities.  We then calculate the probabilities to observe each pattern, however, now we must set $P(A^{K_A}_r)$, $P(B^{K_B}_s)$ and $P(A^{K_A}_r, B^{K_B}_s)$ equal to zero if $r$ or $s$ contain 1's in time-slots that are closer together than $M_d$.  We then calculate the new probabilities $\widetilde{P}(K_A,K_B)$, $\widetilde{P}(K_A)$ and $\widetilde{P}(K_B)$, together with the new conditional probabilities.  Finally, the conditional probabilities are used in Eq. (\ref{cond_mi_def}) to calculate the conditional mutual information.  The approach is best illustrated by an example.

Suppose we have a detector with dead-time of the order of one time-bin width, {\it i.e.}, $M_d=1$.  This means that it is impossible for Alice to observe two photon measurement patterns such as $1100$ or $0110$.  The probability to observe such patterns must be set to zero, hence $P(A^2_{1100})=0$.  This reduces the number of two photon patterns that Alice can observe from $N(N-1)/2$ to $(N-1)(N-2)/2$.  In general, dead-time reduces the total number of allowed two-photon patterns to $(N-M_d)(N-M_d-1)/2$. 

It is convenient to introduce a function $\Delta_{M_d}(X^{K}_r)$ that is 0 if the pattern $X^{K}_r$ contains 1's that are closer together than $M_d$.  Otherwise, the function returns the value of 1.  For example $\Delta_1(1010)=1$, while $\Delta_1(1100)=0$.  The new probabilities to observe measurement patterns can be expressed in terms of $P(A^{K_A}_r,B^{K_B}_s)$, $P(A^{K_A}_r)$ and $P(B^{K_B}_s)$, the probabilities for the case when there is no dead-time effect.  As an example, consider the new probabilities, denoted by a tilde, for the case when $M_d=1$.  We find that
\begin{eqnarray}
\label{deadtimepr}
&\widetilde{P}(A^{K_A}_r,B^{K_B}_s)=\frac{P(A^{K_A}_r,B^{K_B}_s)\Delta_{1}(A^{K_A}_r)\Delta_{1}(B^{K_B}_s)}{P_{00}^2},\nonumber\\
&\widetilde{P}(A^{K_A}_r)=\frac{P(A^{K_A}_r)\Delta_{1}(A^{K_A}_r)}{(P^A_0)^2},\nonumber\\
&\widetilde{P}(B^{K_B}_s)=\frac{P(B^{K_B}_s)\Delta_{1}(B^{K_B}_s)}{(P^B_0)^2}.
\end{eqnarray}
The reason for dividing by either $(P_{00})^2$ or $[P^{A(B)}_{0}]^2$ is to take into account the fact that, after our two clicks, we cannot detect anything.  This is distinct from not observing a click, which happens with joint probability $P_{00}$ and marginal probabilities $P^A_0$ and $P^B_0$.  We note that Eq. (\ref{deadtimepr}) neglects the effect of obtaining a click within the last time-bin, {\it e.g.}, observing a pattern such as $01001$.  In cases like this, we should only divide $P(A^{K_A}_r)$ by $P^{A(B)}_0$.  This is because we don't see the dead-time for the last detection.  When the frame size is large, the relative probability to observe a click within the last time-bin becomes small.  In this regime, our approximation is very good.

The modification of the above results to the case where $M_d>1$ is straightforward.  If we neglect the effects of the frame edge, then $\widetilde{P}(A^{K_A}_r,B^{K_B}_s)=P(A^{K_A}_r,B^{K_B}_s)\Delta_{M_d}(A^{K_A}_r)\Delta_{M_d}(B^{K_B}_s)/[P_{00}]^{2M_d}$.  The marginal probabilities are $\widetilde{P}(X^K_r)=P(X^K_r)\Delta(X^K_r)/[P^X_0]^{2M_d}$, where $X$ is either $A$ or $B$.  When $M_d$ becomes larger,  relative to the frame size, the approximation may seem dubious.  One can explicitly take account of the edges by changing the probabilities such as $\widetilde{P}(A^{2}_1001)$.  However, as $M_d$ becomes large, this also increases the probability that a detector cannot register photons in the beginning of a frame due to dead-time from a click in the previous frame.  The effect of these two edge effects is to act in opposite ways.  One increases the pattern probabilities, while the other acts to decrease them.  The net effect is that, to some extent, both effects compensate for each other.  It is thus still a good approximation to neglect both edges.


\section{Detector jitter}
\label{secV}
One thing we have omitted so far is the temporal response of the detectors.  In any real detector, there can be a randomly fluctuating delay between a photon being incident on the detector and it firing.  This is very important in time-binned experiments, as it can cause a photon to be registered in the wrong time-bin.  This effect is known as detector jitter \cite{ralph}.  In this section we show how jitter can be included within our model.  For the sake of clarity, we illustrate the approach only for $(1,1)$-frames.  The general method, however, can also be applied to other frame classes.

A simple way of modeling jitter is to calculate a discrete set of `jump' probabilities from the temporal response.  
Mathematically, the temporal response is the probability distribution to register a photon at a time $t$ after it was incident on the detector.\footnote{Often the detector's response is well described by a Gaussian.  The approach we will outline, however, makes no assumption about the form of the continuous probability distribution.}  By integrating over the width of each time-bin, we convert the continuous probability distribution into a discrete set of detection probabilities.  Suppose a photon is generated within the $r$-th time-bin.  Let $J_n$ be the probability that we observe a click within the $(r+n)$-th time slot.  The probability to observe the photon within the correct time-bin is thus $J_0$.  Clearly, $\sum_n{J_n}=1$.  Often, $J_n$ is non-zero only for $n=1$ or 2.

One difficulty in modeling jitter is the presence of dark counts.  The single time-bin detection probabilities include dark counts, which are not subject to jitter.  However, the probability $q$ is the same for each time-bin.\footnote{This will not be true in the presence of after-pulsing.  Nevertheless, this does not affect our results as we adopted an overly cautions approach where after-pulsing is modeled by increasing $q$ for every time-bin.  This means that $q$ is the same for every time-bin in our mathematical formalism.}  The dark count probability is thus invariant to shifts in time.  This suggests that contribution from dark counts within ${P}_{ij}$ should also be approximately invariant to temporal shifts.  Hence, to an excellent approximation, it is not effected by jitter.\footnote{This approximation is very good for $q$ small.  To be more quantitative, if the temporal width of each time-bin is 1ns, then the approximation is excellent provided the dark count rate is less than about $10^8$ counts per second.}  This observation means we can use the jump probabilities directly with the probabilities ${P}_{ij}$, without having to separate out the contribution from dark counts.

We begin by looking at the marginal probability for Alice to observe a particular measurement pattern.  It is convenient to modify our notation.  As we are interested in the case where $K_A=K_B=1$, we represent Alice's pattern as $A_i$, where $i$ is the location of the time-bin where the photons are detected.  For example, $A_1$ represents the pattern where Alice observes a click within the first  time-bin.  

As a further simplification, we limit our analysis to the case when the detector's response is short enough so that only $J_0$ and $J_1$ are greater than zero and $J_0+J_1=1$ while $J_2=0$.  Appendix A explains how to generalize the results to the case when $J_2\ne 0$.  Consider two time bins that are away from the edges of the frame.  In these time-bins we observe a single click in the second of them, {\it i.e.}, our measurement pattern is $01$.  If there were no jitter, then this pattern occurs with probability $P^A_0P^A_c$.  When the detector jitter is not negligible, then the pattern  could have arisen from two possible situations.  First, there was no delay and we observe the photons in the correct time-slot, which occurs with probability $J_0$.  Alternatively, jitter could have caused a photon that was in the first time-bin to be registered within the second.  We find that the total probability to observe the pattern 01 is
\begin{equation}
\label{mid1}
\mathcal{P}_1=J_0P^A_cP^A_0+J_1P^A_c.
\end{equation}
If a photon is incident on the last time-bin within a frame, then jitter can cause it to be lost to the frame.  Similarly, a click within the first time-slot of each frame could have come from the previous frame.  Let $\mathcal{P}_e$ be the probability to {\it not} see a click within the last time-bin of a frame.  We find that 
\begin{equation}
\label{end1}
\mathcal{P}_{e}=P^A_0+J_1P^A_c.
\end{equation}
We thus see that the probabilities for Alice to observe a given pattern is 
\begin{eqnarray}
P(A_1)&=&\mathcal{P}_1\mathcal{P}_e(P^A_0)^{N-2},\nonumber\\
P(A_i)&=&\mathcal{P}_1\mathcal{P}_e(P^A_0)^{N-3},\;
\nonumber\\P(A_N)&=&\mathcal{P}_1(P^A_0)^{N-2},
\end{eqnarray}
where $1<i<N$.  The results for Bob will have the same form, but with each $A$ changed to a $B$.

The joint probabilities, $P(A_i,B_j)$, are more involved.  To simplify our exposition, we assume symmetric channel losses, {\it i.e.}, $\eta_A=\eta_B$.  The case where $\eta_A\ne \eta_B$ is described in Appendix B.  Each pattern can be broken up into a small set of events, from which each pattern can be constructed.  For example, one event is described by the probability for both Alice and Bob to observe clicks within the same time-bin.  The probabilities of these events can be expressed in terms of $J_0$, $J_1$ and the single time-bin probabilities $P_{ij}$.  The probabilities $P(A_i,B_j)$ are expressed in terms of the event probabilities.

Let $\mathcal{P}_{11}$ be the probability for Alice and Bob to both observe clicks within the same time-bin.  The fact that the detectors suffer from jitter means that we must consider {\it two} time-bins to calculate $\mathcal{P}_{11}$.  It is found that
\begin{eqnarray}
\label{p11c}
&&\mathcal{P}_{11}=J_0^2 P_{00}P_{cc}\nonumber\\
&&+2J_0J_1 (P_{0c})^2+J_1^2 P_{cc},
\end{eqnarray}
where we use the fact that $P^{AB}_{c0}=P^{AB}_{0c}$ when $\eta_A=\eta_B$.  The probability that both Alice and Bob do not see a click in the last time-bin of their frame is
\begin{equation}
\mathcal{P}^e_{00}=P_{00}+J_1^2 P_{cc}+2J_1 P_{0c},
\end{equation}
where we have again use the fact that $P^{AB}_{0c}=P^{AB}_{c0}$.  

It is possible that Alice and Bob can see clicks in adjacent time-bins.  For example, Alice's detector could fire within the $n$-th time-bin, while Bob's fires within the $(n+1)$-th.  Let $\mathcal{P}_{1*}$ be the probability for Alice's detector to fire in a time-bin directly before Bob's.   Similarly, let $\mathcal{P}_{*1}$ be the probability for Bob to observe a detection in the $n$-th time-bin while Alice sees one in the $(n+1)$-th time-slot.  To calculate $\mathcal{P}_{*1}$ and $\mathcal{P}_{1*}$, we need to consider three time-bins.  We find that
\begin{eqnarray}
\mathcal{P}_{*1}=\mathcal{P}_{1*}=J_0^2 P_{00}(P_{c0})^2\nonumber\\
+J_1^2 P_c P_0 P_{c0}+J_0J_1 P_0\Big[P_{00} P_{cc}+(P_{0c})^2\Big],
\end{eqnarray}
where $P_0=P^A_0=P^B_0$ and $P_c=P^A_c=P^B_c$.  The final situation that we consider is when Alice and Bob obtain clicks in different time-bins, which are not adjacent.  The fact that $J_2=0$ implies that we can be certain that any photons detected by Alice and Bob were not from the same photon pair.  Let $\mathcal{P}_{10}$ be the probability for Alice to observe a click in a time bin when Bob does not see a click in the same or adjacent time bins.  Similarly, $\mathcal{P}_{01}$ is the probability for Bob to observe a click while Alice does not see one in nearby time slots.  We find that 
\begin{eqnarray}
\mathcal{P}_{10}&=&\mathcal{P}_{01}=J_0 P_{00} P_{c0}+J_1 P_{c0} P_0\nonumber\\
&+&J_0 J_1 P_{00} P_{cc}+J_1^2 P_{c0} P_c.
\end{eqnarray}
These event probabilities can be used to construct $P(A_i,B_j)$.

One complication with calculating $P(A_i,B_j)$ is that we be must careful of detection events near the edges of the frame.  For instance, the probability $P(A_1,B_1)$ is different from $P(A_3,B_3)$.  This difference is due to the fact that a detection in the first time-bin could have come from the previous frame.  When the size of the frame becomes large, the relative effects of the edges becomes small.  One could thus neglect the effects of the edges\footnote{We are not neglecting all of the effects of the edges.  We still including the probability that one could loss a photon from the edges of a frame.}.  In this case, the probability for Alice and Bob to observe particular measurement patterns is given by
\begin{eqnarray}
\label{jpattern}
P(A_i,B_i)&=&\mathcal{P}_{11}\mathcal{P}^e_{00}\big[P^{AB}_{00}\big]^{N-3},\nonumber\\
P(A_i,B_{i+1})&=&P(A_{i+1},B_i)=\mathcal{P}_{10}\mathcal{P}^e_{00}\big[P^{AB}_{00}\big]^{N-4},\nonumber\\
P(A_i,B_j)&=&[\mathcal{P}_{1*}]^2\mathcal{P}^e_{00}\big[P^{AB}_{00}\big]^{N-5},\;|i-j|>1.\nonumber\\
\end{eqnarray}
In Appendix C, we give the full form of $P(A_i,B_j)$, where edge effects are not neglected.  Using the probabilities (\ref{jpattern}), we find that
\begin{eqnarray}
\label{p1jitter}
&&P(K_A=1,K_B=1)=\mathcal{P}^e_{00}(P^{AB}_{00})^{N-5}\Big[N\mathcal{P}_{11}(P^{AB}_{00})^2\nonumber\\
&&+2(N-1)\mathcal{P}_{10}P^{AB}_{00}+(N-1)(N-2)\mathcal{P}_{1*}^2\Big].
\end{eqnarray}
The post-selected information per detected photon pair is thus
\begin{eqnarray}
\label{jitterHpost}
& H_{d}(A:B|K_A=1,K_B=1)=\frac{1}{N(\eta^2\lambda+q^2)}\nonumber\\
&\times \Big[P(K_A=1,K_B=1)\nonumber\\
& \times\{2\log_2 N-\log_2 P(K_A=1,K_B=1)\}\nonumber\\
& +NP(A_i,B_i)\log_2 P(A_i,B_i)\nonumber\\
& +2(N-1) P(A_i,B_{i+1})\log_2 P(A_i,B_{i+1})\nonumber\\
& +(N-1)(N-2)P(A_i,B_j)\log_2 P(A_i,B_j)\Big].
\end{eqnarray}
The assumption that we neglect edge effects means that Eqs. (\ref{jpattern}), (\ref{p1jitter}) and (\ref{jitterHpost}) are all valid only for $N\ge 6$.  In Appendix C, we compare the approximate results given above with the more complicated exact results.  It is shown that even for $N=8$, the difference between the exact and approximate expressions can be very small (less than 0.1\%).  Thus we can safely use the approximate expression given in Eq. (\ref{jitterHpost}).

\section{Results for a mode-locked laser pumping a SPDC source}
The previous results will now be illustrated by looking at a specific experimental setup.  The situation we consider is a mode-locked laser that produces a train of coherent pulses that pump a nonlinear crystal.  The pulses are generated such that each pulse is coherent to one another \cite{cqo,cqo2,cqo3}.  We fix the parameters of the crystal and laser such that we observe SPDC that produces a pair of photons that are correlated in time.  The down-converted photon pair is split with one half kept by Alice, while the other is sent to Bob.  The experimental configuration is shown in Fig. \ref{experiment}.  

The spacing of the pulses define natural time-bins for Alice and Bob.  Alice and Bob thus choose the widths of their time-bins so that they contain a single pulse.  To a good approximation, the probability that Alice and Bob observe $m$ photon pairs in each time-bin is given by a Poissonian distribution
\begin{equation}
\label{poissonian}
P_s(m)=e^{-\lambda}\frac{\lambda^m}{m!},
\end{equation}
where $\lambda$ is the average number of photon pairs generated in each time-bin.  For this source, the moment generating function, defined in Eq. (\ref{momentgf}), is just  $M(\nu,\xi)=\exp(\lambda[-\eta_A\nu-\eta_B\xi+\eta_A\eta_B\nu\xi])$.  Using this within Eqs. (\ref{piprobs}) and (\ref{fullprobs}) yields joint detection probabilities for each time-bin
\begin{eqnarray}
\label{jdp}
P_{00}&=&(1-q)^2 e^{\left(-\lambda[\eta_A+\eta_B-\eta_A\eta_B]\right)},\\
P_{0c}&=&(1-q)e^{-\lambda\eta_A}-(1-q)^2 e^{-\lambda[\eta_A+\eta_B-\eta_A\eta_B]},\nonumber\\
P_{c0}&=&(1-q)e^{-\lambda\eta_B}-(1-q)^2 e^{-\lambda[\eta_A+\eta_B-\eta_A\eta_B]},\nonumber\\
P_{cc}&=&1-(1-q)\left[e^{-\lambda\eta_A}+e^{-\lambda\eta_B}\right]\nonumber\\
&+&(1-q)^2 e^{(-\lambda[\eta_A+\eta_B-\eta_A\eta_B]}.\nonumber
\end{eqnarray}
The marginal probabilities for Alice (Bob) are $P^{A(B)}_0=(1-q)e^{-\lambda\eta_{A(B)}}$ and $P^{A(B)}_c=1-P^{A(B)}_0$.

Suppose that the main sources of errors are losses and dark counts.  One can use Eq. (\ref{jdp}) directly within (\ref{postinfophoton}) to determine how many shared bits per detected photon we can extract using only $(1,1)$-frames.  The extra information contained in $(2,2)$-frames can be calculated using Eq (\ref{post2}).  These results can be used to optimize the frame size $N$.  Furthermore, one can also investigate how the experimental parameters affect the number of bits per photon.  This could be important, for instance, in evaluating the advantages of improving the detector's efficiency.  

To illustrate our results we look at typical parameters for two detectors: a single-photon avalanche detector (SPAD) and a superconducting nanonwire detector.  We assume that we have time-bins of width 130 ps.  The SPAD has efficiency $\eta=0.7$, dark count rate of 500/s and an after-pulsing rate of 0.5\%.  The effective dark count probability, which includes the effects of after-pulsing, is $q=6.53\times 10^{-8}$.  For the superconducting nanowire detectors $\eta=0.9$, the dark count rate is 1/s and the after-pulsing rate effectively zero.  We calculate the dark count probability as $q=1.3\times 10^{-10}$.  Figures 2 and 3 shows the information within $(1,1)$ and $(2,2)$-frames as a function of the frame size $N$, for a SPAD and superconducting nanowire detector, for two different values of $\lambda$. 

We see in Fig. 2 that, for $N=1000$, $(2,2)$-frames can contain a significant fraction of the total shared bits.  This is not always true, however, as shown in Fig. 3.  Another important point to note from Fig. 3 is that, for $N=3000$, we can extract over 11 bits per photon pair using either of the two detectors.  For both the detectors we considered, the efficiency is high.  If the detectors have low efficiencies, then we would obtain less information; it would thus become crucial to optimize the frame size.  For example, consider a detector with $q=6.53\times 10^{-8}$ and $\eta=0.3$.  We find that, for a source with $\lambda=5.33\times 10^{-5}$,  we obtain 10.3 bits from the $(1,1)$-frames by choosing $N=3579$.

\begin{figure}
\center{\includegraphics[width=7cm,height=!]
{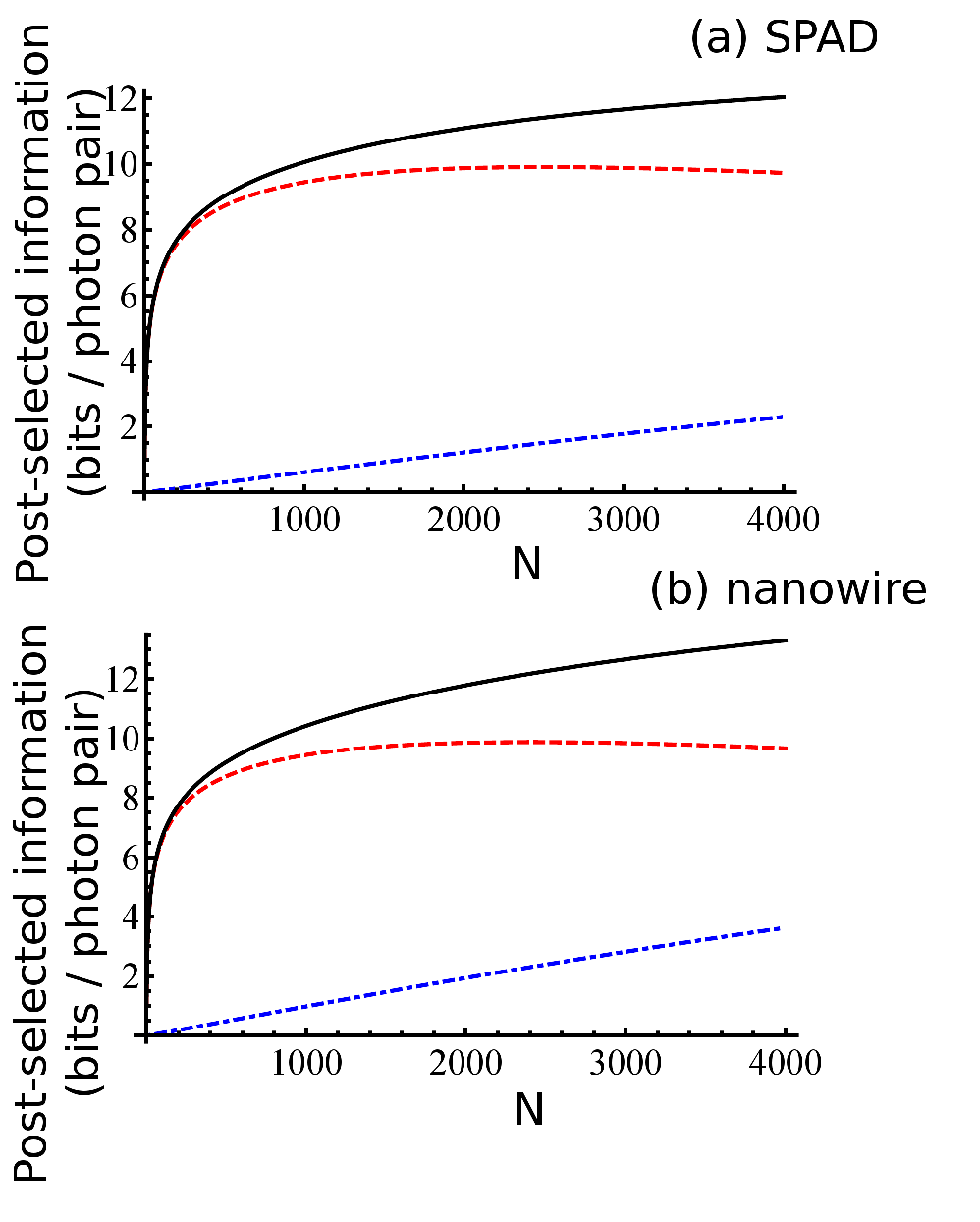}}
\caption{Shared bits per detected photon pair as a function of the frame size $N$.  All plots are for $\lambda=5.33\times 10^{-5}$.  In both (a) and (b), the dashed red line is for $(1,1)$-frames, the dotted blue line is for $(2,2)$-frames and the solid black lines if for both the $(1,1)$ and $(2,2)$-frames.  Fig. (a) is for a single photon avalanche detector with $\eta=0.7$ and $q=6.53\times 10^{-8}$.  Fig. (b) is for a nanowire detector with $\eta=0.9$ and $q=1.3\times 10^{-10}$ (color online).}
\label{p2}
\end{figure}

\begin{figure}
\center{\includegraphics[width=7cm,height=!]
{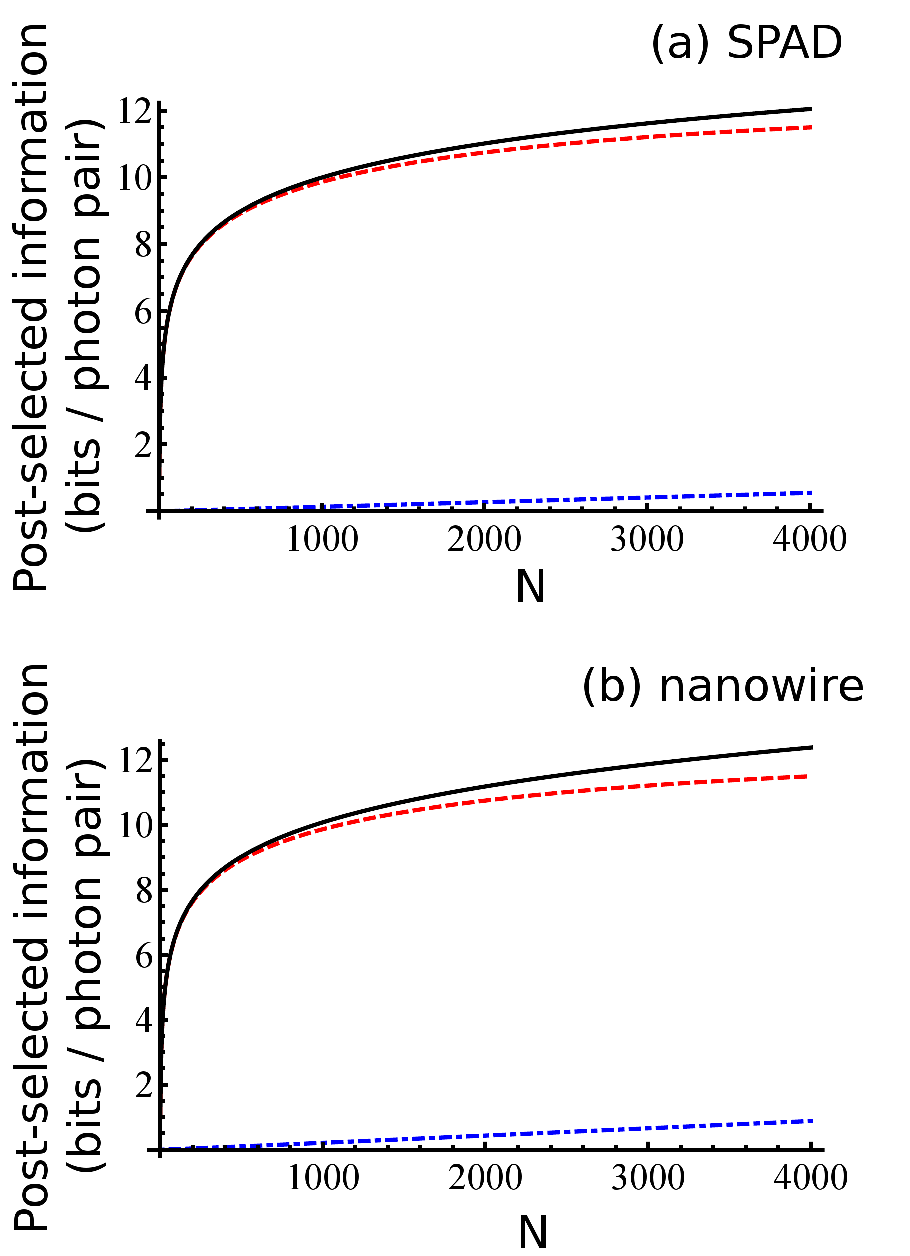}}
\caption{Shared bits per detected photon pair as a function of the frame size $N$.  All plots are for $\lambda=1.0\times 10^{-5}$.  In both (a) and (b) the dashed red line is for $(1,1)$-frames, the dotted blue line is for $(2,2)$-frames and the solid black lines if for both the $(1,1)$ and $(2,2)$-frames.  Fig. (a) is for a single photon avalanche detector with $\eta=0.7$ and $q=6.53\times 10^{-8}$.  Fig. (b) is for a nanowire detector with $\eta=0.9$ and $q=1.3\times 10^{-10}$ (color online).}
\label{p3}
\end{figure}

The design of error correcting codes for $(2,2)$-frames can be difficult.  If we find that, for given values of loss and the dark count rate, $H_{d}(A:B|K_A=2,K_B=2)$ is negligible, then we know that it would not be worth using these frames.\footnote{This conclusion still holds if jitter is significant.  This is because additional errors cannot increase the shared information.}  This result also provides a good guide to determine the regime that one must work in so that $(2,2)$-frames contribute significantly.  Similarly, one can use the results of Sec. III to calculate the information within $(2,1)$-frames.  One could thus investigate the gains from developing error correcting codes for these and other situations.

\begin{figure}
\center{\includegraphics[width=7cm,height=!]
{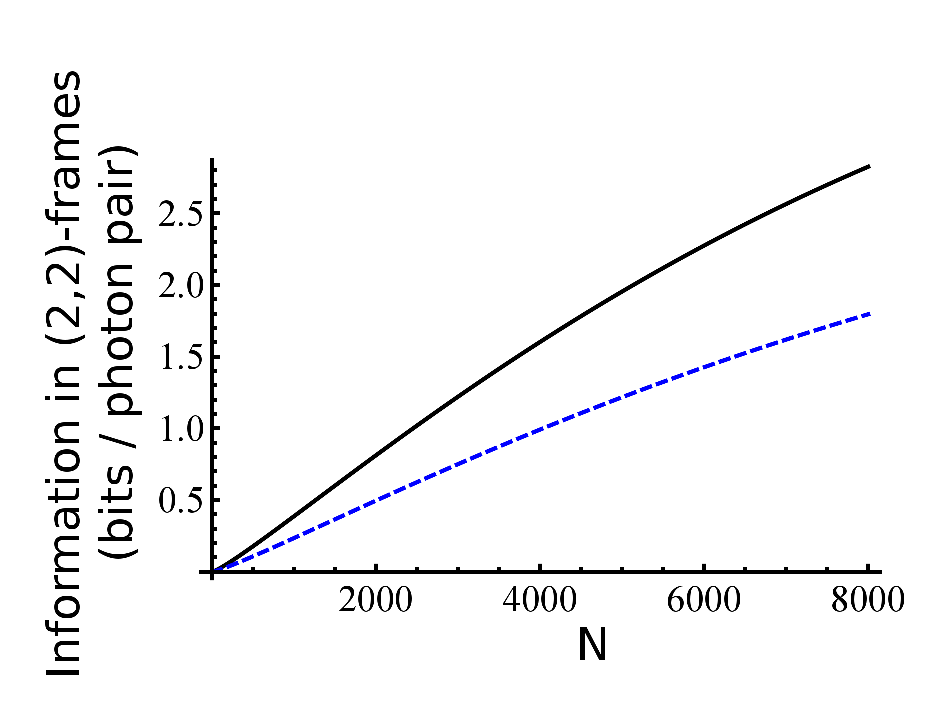}}
\caption{Shared bits per detected photon pair as a function of the frame size $N$.  Both plots are for $(2,2)$-frames with $\lambda=5.33\times 10^{-5}$.  The black curve is for a superconducting nanowire detector with $\eta=0.9$ and $q=1.3\times 10^{-10}$, while the dashed blue curve is for a SPAD with $\eta=0.7$ and $q=6.53\times 10^{-8}$.  The dead-time for the superconducting nanowire corresponded to 154 time-bins, while the dead-time for the SPAD corresponded to 230 time-bins (color online).}
\label{p3dt}
\end{figure}

The previous results did not include the effects of detector dead-times.  However, if we are to fully evaluate the information contained within (2,2)-frames, then we must take this effect into account.  Results for this can be found using the approach detailed in Sec. \ref{secIV}.  For the SPAD, the dead-time is 30 ns, which corresponds to approximately 230 time-bins, while for the superconducting nanowire, the dead-time is 20 ns, which corresponds to 154 time-bins.  Figure \ref{p3dt} compares the the shared information in (2,2)-frames for the case of dead-time and no dead-time, where part (a) is for the SPAD and (b) is for the superconducting nanowire.  Both curves are for $\lambda=5.33\times 10^{-5}$.  While dead-time can reduce the information, we still see that useful information can still be extracted from (2,2)-frames. 

In many realistic situations, detector jitter is non-negligible.  We can include the effects of jitter by using the formalism described in Sec. 5.  There is, however, a subtlety when one applies the theoretical results to an experiment.  It is common to calculate the heralded efficiency directly from experimental data.  Detector jitter decreases the probability to observe photons within a particular period of time.  If one is not careful, then we could over estimate losses and hence under estimate $\eta$.  To illustrate this, consider the example where we have a source that can produce a single photon within a specific time-bin.  One can use this source to estimate $\eta$ by looking at the probability $w$ to detect the photon.  If our detectors suffer from jitter, then $w\ne \eta$.  Instead, we have $w=\eta J_0$ when we neglect dark counts. 

In some situations, under estimating the efficiency can be a good thing.  For example, a reduction in $\eta$ will decrease our estimate of the number of  bits we can extract.  We could use this as a crude way of taking account of jitter.  Such an approach would, however, be too pessimistic if we have already included jitter explicitly within our model.  In the rest of this section, we will assume that the total efficiency $\eta$ has been estimated such that it is completely associated with losses.

Jitter causes a decrease in the correlation within Alice and Bob's timing information.  This inevitably leads to a decrease in the number of shared bits.  To evaluate the effects we calculate $H_d(A:B|K_A=1,K_B=1)$ for the SPAD and superconducting nanowire detector, with a $\lambda= 2.0\times 10^{-5}$.  We take $J_0=0.9$ for the SPAD and $J_0=0.97$ for the nanowire detector.  Figure \ref{p3} shows $H_d(A:B|K_A=1,K_B=1)$ plotted as a function of $N$.  The solid black curve corresponds to the SPAD, while the dashed blue line is for the superconducting nanowire.  We see that for an appropriate choice of $N$, we can still obtain greater than 10 bits per photon using either detector.  One can get a better feel for how jitter affects use by looking at how $H(A:B|K_A=1,K_B=1)$ changes with $J_0$.  Consider a setup with $\lambda=2.0\times 10^{-5}$, $\eta=0.7$ and $q=6.53\times 10^{-8}$, i.e. the parameters for the SPAD.  We could extract 11.1 bits per photon for $N=4000$, if we had no jitter ($J_0=1$).  If instead, $J_0=0.9$, then we could extract 10.2 bits per photon for frames of size $N=4000$.


\begin{figure}
\center{\includegraphics[width=7cm,height=!]
{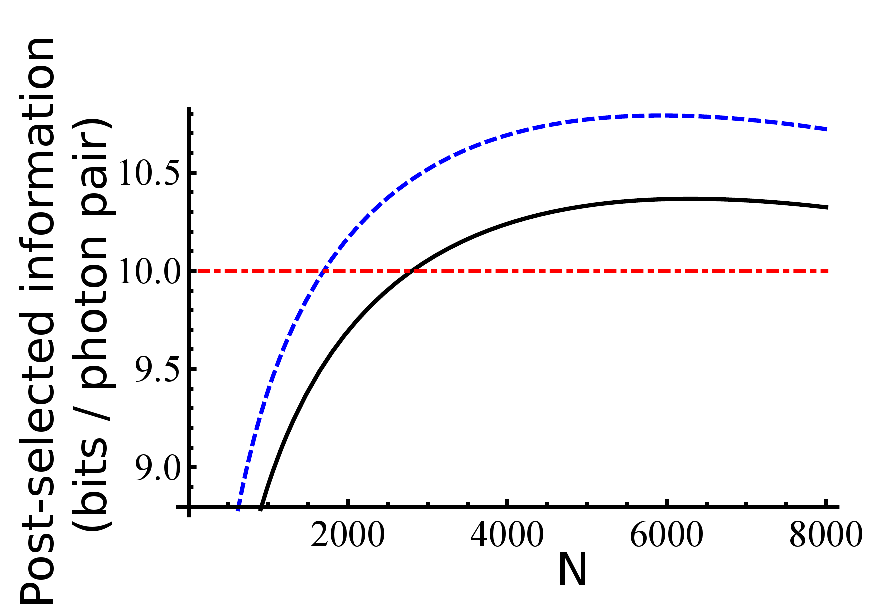}}
\caption{Plot of shared bits per detected photon pair as a function of the frame size $N$.  All plots are for $(1,1)$-frames with $\lambda= 2.0\times 10^{-5}$.  The solid black line corresponds to the SPAD with $\eta=0.7$, $q=6.53\times 10^{-8}$ and $J_0=0.9$.  The dashed blue line is for the superconducting nanowire with $\eta=0.9$, $q=1.3\times 10^{-10}$ and  $J_0=0.97$.  Finally, the dotted and dashed red line corresponds to the threshold for 10 bits per photon pair (color online). }
\label{p3}
\end{figure}

\section{Conclusions}
The time-of-arrival degree of freedom provides an experimentally viable means of implementing high-dimensional quantum information protocols, and is particular well suited for quantum communication.  One important example of this is high-dimensional QKD.   Such schemes can, however, be hampered by the difficult in performing error correction.  A pertinent example of this is in time-bin based QKD, where unlike in polarization based QKD, one cannot use each photon's arrival time to help correct errors.  Instead, it is common to split the arrival time into discrete time-bins, which are grouped together to form frames.  A key question is how this affects the amount of shared information that Alice and Bob can extract.  We answer this question and obtain general results for the maximum number of extractable shared bits for photons entangled within their time-of-arrival, when using frame encoding.


Our results go beyond existing work in a number of areas.  First, we present results for frames that contain photon pairs within multiple time-bins.  We can thus investigate how many bits are lost by neglecting such events and when such events should be kept.  The results can thus be used to improve the efficiency of extraction of shared bits from noisy experimental setups.  

Another way in which the current work improves on existing works is in the range of errors considered.  We study systems that suffer from asymmetric losses, dark counts, after-pulsing, dead time and jitter.  The formalism also works for a general choice of source.  
The results for the case of jitter are of particular interest.  We have found analytic expressions for the extractable information when we have jitter in addition to losses and dark counts.  This could be important for optimization of high-dimensional QKD.  For example, in some experiments, we have freedom in the choice of the time bin width.  Often, one chooses the width such that it minimizes jitter.  By using our results, we can more efficiently choose the time-bin width so as to optimize the shared information.

The results are illustrated by considering at entangled photons generated by a nonlinear crystal that is pumped by a mode-locked laser.  This source produces a train of pulses that are temporally coherent to one another.  Two different types of detector were considered, a SPAD and a superconducting nanowire detector.  The results show that under appropriate conditions, we can chose a frame size so as to extract over 10 shared bits per photon pair.

One issue we have not considered is how one might actually extract the shared information, {\it i.e.}, reconciliation.  There has been some work in this area \cite{Kochman,wornell,wornell2}.  Some of the present authors have developed a reconciliation protocol that is tailored to the case of a mode-locked laser pumping a down-conversion source \cite{kumor}.  This protocol can treat multi-photon events and can recover some of the frame-to-frame information contained within the photon number uncertainty.  

\section*{Acknowledgements}
We would like to thank Paul Kwiat, Bradley Christensen and Daniel Kumor for useful discussions.  This research was supported by the DARPA InPho program through the US Army Research Office award W911NF-10-0395.  DJG gratefully acknowledges the financial support of the Office of Naval Research MURI $\#$N00014-13-0627.

\section*{Authors contribution statement}
S. M. Barnett and D. J. Gauthier contributed to the development of the problem.  All authors contributed equally in development of the most useful quantities to calculate for time-bin encoded experimental applications.  T. Brougham, with feedback from S. M. Barnett, C. F. Wildfeuer and especially D. J. Gauthier, carried out calculations for all plots and developed the model for treating jitter.

\appendix

\section{Modeling jitter for a long-tailed detector response}
The calculations in Sec. 5 assumed that $J_n=0$ for $n\ge 2$.  This is consistent with the detector's temporal response ({\it i.e.}, its probability distribution) being effectively zero over more than two time-bins.  While this assumption is often true, there are detectors for which it would not hold.  In this Appendix, we briefly outline how to generalize the previous results.  The aim is not to present extensive results, but instead to show how to adapt the previous results.  The general approach is illustrated by investigating at the case when $J_2\ne 0$, but $J_3,J_4,...=0$, hence $J_0+J_1+J_2=1$.  

Recall, $J_0$ is the conditional probability to register a photon in the correct time-bin, {\it i.e.}, the time-bin in which the photon actually was incident on the detector.  The conditional probability to register a click $n$ time-bins after it was incident on the detector, is given by $J_n$.  We first consider Alice's (or equivalently Bob's) marginal probabilities.  In the absence of jitter, the probability for Alice to see a click in a given time bin is $P^A_c$.  When we do have jitter, then a detected photon could have originated in previous time slots.  The probability to observe a click thus changes.  To calculate the new probability $\mathcal{P}_1$, we consider three time-bins, as the term $J_2$ can cause a photon to jump over two time-bins.  We find that the probability for Alice to see a click in a given time bin is 
\begin{equation}
\label{mid2}
\mathcal{P}_1=P^A_c\left[J_2+J_1 P^A_0 + J_0(P^A_0)^2\right].
\end{equation}
where $P^A_0=1-P^A_c$.  Equation (\ref{mid2}) is composed of three separate terms.  The first term in Eq. (\ref{mid2}) corresponds to a photon that has `jumped' two time-bins due to jitter.  The second term is for a photon that is detected within a time-bin directly after the correct one.  Finally, the third terms corresponds to the detector firing within the time slot in which it was incident on the detector.  

To calculate the probability $\mathcal{P}_e$, that we don't see a click at the last time-bin of a frame we must consider two time-bins.  We find that the probability is 
\begin{equation}
\label{end2}
\mathcal{P}_{e}=J_2[P^A_c]+J_1[P^A_0P^A_c]+[P^A_0]^2.
\end{equation}
Again, we have three terms corresponding to three possible ways in which the event could be realized.  The probability to observe a given measurement pattern is again constructed from $\mathcal{P}_1$, $\mathcal{P}_e$ and $P^A_0$.  One thing we must take care of are photons detected in time slots near the edge of each frame.  It is possible that these correspond to photons from previous frames, which are registered in a later frame due to jitter.  These edge effects mean that $P(A_1)$ or $P(A_2)$ will not equal $P(A_i)$, where $i$ is a time bin in the middle of the frame.  Similarly, $P(A_N)\ne P(A_i)$, where again $i$ corresponds to a time bin near the middle of the frame.  We find that
\begin{eqnarray}
P(A_1)&=&\mathcal{P}_1\mathcal{P}_{e} (P^A_0)^{N-3}, \nonumber\\
P(A_2)&=&\mathcal{P}_1\mathcal{P}_{e} (P^A_0)^{N-4},\nonumber\\
P(A_i)&=&\mathcal{P}_1\mathcal{P}_{e}(P^A_0)^{N-5},\nonumber\\
P(A_{N-1})&=&\mathcal{P}_1\mathcal{P}_{e}(P^A_0)^{N-4},\nonumber\\
P(A_N)&=&\mathcal{P}_1 (P^A_0)^{N-3},
\end{eqnarray}
where $i<2<N-1$.  The probability for Alice to post-select on a $K_A=1$ frame is $P(K_A=1)=\sum_j{P(A_j)}$.  The results for Bob will have the same form.

The joint probabilities $P(A_m,B_n)$ can be calculated in the same fashion by first recalculating $\mathcal{P}_{11}$, $\mathcal{P}^e_{00}$, $\mathcal{P}_{10}$ and $\mathcal{P}_{1*}$.  However, now we require an extra term $\mathcal{P}_{1**}$, which corresponds to Alice seeing a photon two time-bins before Bob does.  Each of these probabilities will again be calculated by looking at several time-bins.  For instance, to calculate $\mathcal{P}_{11}$, we must consider three time-bins for Alice and Bob.  As an example, the new form for $\mathcal{P}_{11}$ is 
\begin{eqnarray}
\label{appA}
\mathcal{P}_{11}&=&J_0^2P_{00}^2P_{cc}+J_1^2P_{00}P_{cc}+J_2^2P_{cc}\nonumber\\
&+&2J_0J_1 P_{00}P_{c0}P_{cc}+2J_0J_2 P_{c0}P_0P_1 \nonumber\\
&+&2J_1J_2 P_{c0}P_1,
\end{eqnarray}
where we have used the fact that $P_{c0}=P_{0c}$ when $\eta_A=\eta_B$.  Notice that (\ref{appA}) contains more terms than Eq. (\ref{p11c}),  which was derived for $J_2=0$.  These extra terms result from the fact that now the detector's response is longer and thus jitter can cause a photon to be register two time slots after it was incident on the detector.

\section{Jitter with asymmetric losses}
The results for jitter given in Sec. \ref{secV} assumed that $\eta_A=\eta_B$, to simplify the expressions.  In this Appendix, we briefly show how the results are modified for asymmetric loss.  The marginal probabilities for Alice and Bob contain terms that depend only on $\eta_A$ or on $\eta_B$.  Thus, there is no need to modify these results.  The joint probabilities $P(A_i,B_j)$ will, however, need to be modified.  

The first step is to calculate the probabilities for the individual events, {\it e.g.} $\mathcal{P}_{11}$, $\mathcal{P}^e_{00}$, etc.  The key issue is that now $P_{c0}\ne P_{0c}$, which was implicitly assumed within the derivations.  As a first step, consider the probability that Alice and Bob both see a click within the same time-bin.  We find that 
\begin{equation}
\mathcal{P}_{11}=J_0^2 P_{00}P_{cc}+2J_0J_1 (P_{0c}P_{c0})+J_1^2 P_{cc}.
\end{equation}
The new probability that both Alice and Bob don't detect photons in the last time-bin of their frame is
\begin{equation}
\mathcal{P}^e_{00}=P_{00}+J_1^2 P_{cc}+J_1[P_{0c}+P_{c0}].
\end{equation}
The probability for Alice and Bob to obtain clicks in adjacent time-bins is given by $\mathcal{P}_{1*}$ and $\mathcal{P}_{*1}$.  Previously, we found that $\mathcal{P}_{1*}=\mathcal{P}_{*1}$, which is not true in general.  We find that 
\begin{eqnarray}
\mathcal{P}_{1*}&=&J_0^2 P_{00}P_{c0}P_{0c}+J_1^2 P^B_c P^A_0 P_{c0}\nonumber\\
&+&J_0J_1 \Big[P^A_0P_{00} P_{cc}+P^B_0P_{0c}P_{c0}\Big],\nonumber\\
\mathcal{P}_{*1}&=&J_0^2 P_{00}P_{c0}P_{0c}+J_1^2 P^A_c P^B_0 P_{0c}\nonumber\\
&+&J_0J_1 \Big[P^B_0P_{00} P_{cc}+P^A_0P_{0c}P_{c0}\Big].
\end{eqnarray}
The final event probability is for the case when Alice and Bob obtain clicks in different and non-adjacent time-bins.  Again we will find that $\mathcal{P}_{10}\ne \mathcal{P}_{01}$, given explicitly by
\begin{eqnarray}
\mathcal{P}_{10}&=&J_0 P_{00} P_{c0}+J_1 P_{c0} P^B_0\nonumber\\
&+&J_0 J_1 P_{00} P_{cc}+J_1^2 P_{c0} P^B_c.\nonumber\\
\mathcal{P}_{01}&=&J_0 P_{00} P_{0c}+J_1 P_{0c} P^A_0\nonumber\\
&+&J_0 J_1 P_{00} P_{cc}+J_1^2 P_{0c} P^A_c.
\end{eqnarray}
The event probabilities will, again, be used to construct the joint frame probabilities $P(A_i,B_j)$.

As before, we simplify our results by assuming that we can neglect edge effects.  Using this assumption, we find that 
\begin{eqnarray}
P(A_i,B_i)&=&\mathcal{P}_{11}\mathcal{P}^e_{00}\big(P^{AB}_{00}\big)^{N-3},\nonumber\\
P(A_i,B_{i+1})&=&\mathcal{P}_{1*}\mathcal{P}^e_{00}\big(P^{AB}_{00}\big)^{N-4},\nonumber\\
P(A_{i+1},B_i)&=&\mathcal{P}_{*1}\mathcal{P}^e_{00}\big(P^{AB}_{00}\big)^{N-4},\nonumber\\
P(A_i,B_j)&=&\mathcal{P}_{10}\mathcal{P}_{01}\mathcal{P}^e_{00}\big(P^{AB}_{00}\big)^{N-5},
\end{eqnarray}
where $|i-j|>1$.  The conditional probabilities and all the relevant entropic quantities can now be calculated as before.

\section{Comparison of the exact and the approximate results for jitter}
In this Appendix, we compare the approximate results for jitter to the longer, but more accurate results.  In all of what follows, we assume that only $J_0$ and $J_1$ are not equal to zero and that $\eta_A=\eta_B$.

In general, the frame edges influence the probabilities for each pattern, e.g. $P(A_1,B_j)\ne P(A_3,B_j)$.  This is because we analysis each frame separately.  We thus loose information about what happens in the time-bins directly before the beginning of each frame.  The probabilities shown in Eq. (\ref{jpattern}) are derived by neglecting the edges.  When we include the edges, we find that the probabilities become 
\begin{widetext}
\begin{eqnarray}
\label{fulljitterp}
P(A_1,B_1)&=&\mathcal{P}_{11}\mathcal{P}^e_{00}\big(P^{AB}_{00}\big)^{N-2},\nonumber\\
P(A_i,B_i)&=&\mathcal{P}_{11}\mathcal{P}^e_{00}\big(P^{AB}_{00}\big)^{N-3},\;i=2,...,N-1,\nonumber\\
P(A_N,B_N)&=&\mathcal{P}_{11}\big(P^{AB}_{00}\big)^{N-2},\nonumber\\
P(A_1,B_2)&=&P(A_2,B_1)=\mathcal{P}_{10}\mathcal{P}^e_{00}\big(P^{AB}_{00}\big)^{N-3},\nonumber\\
P(A_1,B_j)&=&P(A_j,B_1)=(\mathcal{P}_{1*})^2\mathcal{P}^e_{00}\big(P^{AB}_{00}\big)^{N-4},\;1<j<N,\nonumber\\
P(A_1,B_N)&=&P(A_N,B_1)=(\mathcal{P}_{1*})^2\big(P^{AB}_{00}\big)^{N-3},\nonumber\\
P(A_i,B_{i+1})&=&P(A_{i+1},B_i)=\mathcal{P}_{10}\mathcal{P}^e_{00}\big(P^{AB}_{00}\big)^{N-4},1<i<N-1\nonumber\\
P(A_i,B_N)&=&P(A_N,B_i)=(\mathcal{P}_{1*})^2\big(P^{AB}_{00}\big)^{N-4},\;1<i<N,\nonumber\\
P(A_{N-1},B_N)&=&P(A_N,B_{N-1})=\mathcal{P}_{10}\big(P^{AB}_{00}\big)^{N-3},\nonumber\\
P(A_i,B_j)&=&(\mathcal{P}_{1*})^2\mathcal{P}^e_{00}\big(P^{AB}_{00}\big)^{N-5},1<i,j<N.
\end{eqnarray}
\end{widetext}
There will be $N-2$, $P(A_i,B_i)$ terms for $1<i<N$.  Similarly, there are $N-3$, $P(A_1,B_i)$, $P(A_i,B_1)$, $P(A_j,B_N)$ and $P(A_N,B_j)$ terms, where $2<i<N$ and $1<j<N-1$.  One can also verify that there are $N-1$ terms such as $P(A_j,B_{j+1})$ and $P(A_{j+1},B_j)$, where $j=1,...,N-1$.  Finally, the number of remaining terms can be found by recalling that the joint probability contains a total of $N^2$ different outcomes.  

\begin{figure}
\center{\includegraphics[width=7cm,height=!]
{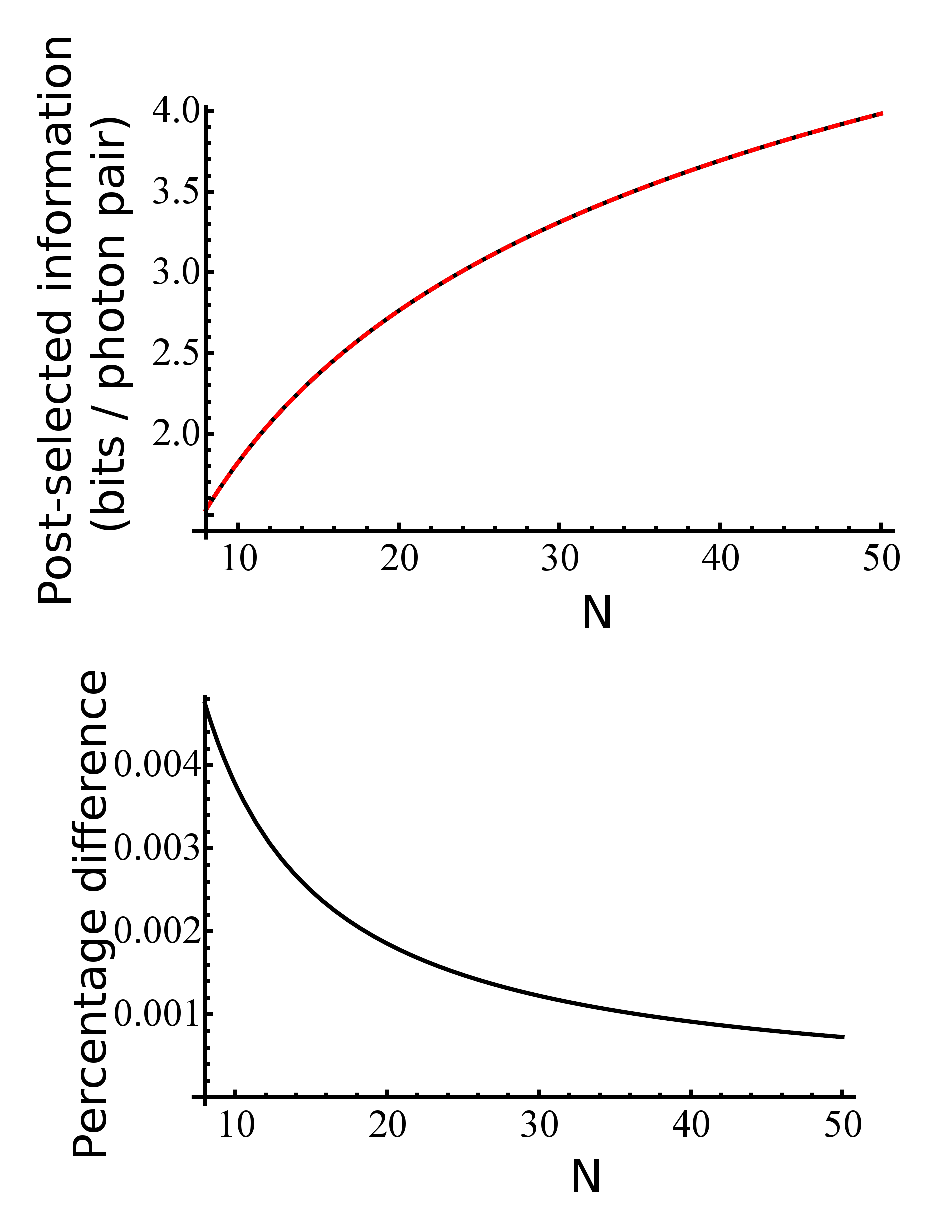}}
\caption{Comparison of the exact expression for $H_d(A:B|K_A=1,K_B=1)$ with the approximate expression (\ref{jitterHpost}).  Fig. (a) shows plots of the post-selected information $H_d(A:B|K_A=1,K_B=1)$, as a function of the frame size $N$.  The black curve is the exact result, while the dashed red curve is the approximate result. Fig. (b) shows a percentage difference between the exact and approximate results, as a function of the frame size.   All curves are for $\eta_A=\eta_B=0.3$, $\lambda=5.33\times 10^{-4}$, $q=3.9\times 10^{-8}$ and $J_1=0.4$ (color online). }
\label{p4}
\end{figure}

Equation (\ref{fulljitterp}) is significantly more complicated than (\ref{jpattern}).  Using these probabilities, we calculate $H_d(A:B|K_A=1,K_B=1)$ and compare this with the approximate result given in (\ref{jitterHpost}).  Figure \ref{p4} (a) shows a direct comparison for $\eta_A=\eta_B=0.3$, $\lambda=5.33\times 10^{-4}$, $q=3.9\times 10^{-8}$ and $J_1=0.4$ as a function of $N$.  The solid black curve is exact expression, while the dashed red curve is the approximate expression. The percentage difference between the exact and approximate results is shown in Fig. \ref{p4} (b).  We see that the agreement between the two results is excellent for large $N$.  Somewhat surprisingly, the approximation is accurate to less that $1\%$ for frames as small as $N=8$.  The match between the exact results and the approximate ones holds also for different values for $\eta_A$, $\eta_B$, $\lambda$ and $q$.  For example, for $\eta_A=\eta_B=0.7$, $\lambda=5.33\times 10^{-5}$, $q=6.53\times 10^{-8}$ and $J_1=0.1$, then we find a percentage difference of less than $0.001\%$ for $N=10$.

\end{document}